# Are electromagnetic phenomena derivable from extended Einstein equations?

Jacob Biemond [*]

*Vrije Universiteit, Amsterdam 1971-1975*
*Fontys University of Professional Education, Eindhoven, The Netherlands*

## Abstract

A new term describing interactions between charge and potentials may be added to the right hand side of Einstein's equations. In the proposed term an additional tensor has been introduced containing a charge density, analogous to the energy-momentum tensor containing a mass density. (The contribution of the electromagnetic fields has not been considered in this work.) The metric components in the new equations may contain charge and mass contributions.

In the special relativistic case a set of four vacuum Maxwell equations is deduced from the new equations containing a special relativistic charge density. A related result was recently found by Jefimenko from a special relativistic transformation. Moreover, from the postulated equations the usual special relativistic electric energy and Lorentz force can be calculated, when the approximated electromagnetic metric is used. In the non-relativistic case the usual Maxwell equations with rest charge density are obtained.

In the presence of a static gravitational field a set of four Maxwell equations is obtained applying in the general relativistic case. From these equations the generally accepted isotropic velocity of light can be deduced. In addition, calculated generalized Maxwell equations show that vacuum permittivity and permeability may differ from unity value in the presence of a big charge $Q$ or a big mass $M$.

## 1. Introduction

Using special relativistic transformations Jefimenko [1] recently generalized the vacuum Maxwell equations. In his equations the usual charge density was replaced by a special relativistic charge density $\rho$ given by

$$\rho \equiv (1 - v^2/c^2)^{-1/2} \rho_0 \tag{1}$$

Here $\rho_0$ is the rest charge density, analogous to the rest mass density in the theory of special relativity, $v$ is the (relativistic) velocity of the charge element and $c$ is the velocity of light.

Landau and Lifshitz [2 (§90)] and Van Bladel [3 (ch. 9)] (see also [4 (§22.4), 5 (§20)]) deduced vacuum Maxwell equations at relativistic velocities and in the presence of a gravitational field. They obtained these equations from a pair of basic relations containing the electromagnetic field tensors $F_{ij}$ and $F^{ij}$. In absence of a gravitational field the generalized Maxwell equations in [2], however, do not yield the special relativistic charge density $\rho$ of (1) but the rest charge density $\rho_0$.

In the presence of a static gravitational field the calculated Maxwell equations in [2,3] predict an isotropic light velocity $c'$ of magnitude

$$c' = (1 - 2GM/c^2R)^{1/2} c \cong (1 - GM/c^2R)c = (1 + c^{-2}\Phi)c \tag{2}$$

Here $G$ is the gravitational constant, $M$ is a big mass, $R$ is the scalar value of $\mathbf{R}$, the radius vector from mass $M$ to a charge $q_0$ associated with a mass $m_0$ and $\Phi = -GM/R$ is the

---

[*] Present address: Sansovinostraat 28, 5624 JX Eindhoven, The Netherlands. E-mail: gravi@gewis.nl

Newtonian potential working on $m_0$. (Throughout this paper it will be assumed that $GM/c^2R \ll 1$.) However, result (2) is not satisfying, since for the isotropic light velocity one expects instead of (2)

$$c' = (1 - 2GM/c^2R)c \qquad (3)$$

In this paper Jefimenko's special relativistic vacuum Maxwell equations and general relativistic Maxwell equations yielding the light velocity of (3) are deduced from a new set of postulated equations, partly analogous to Einstein's gravitational equations. The generalized Maxwell equations show that vacuum permittivity and permeability may differ from unity value in the presence of an additional big charge $Q$ or a big mass $M$. Moreover, in the special relativistic case starting from the electromagnetic metric the total electric energy and the Lorentz force are obtained.

It is noticed that the demonstrated reduction of the proposed equations leading to the non-relativistic Maxwell equations corresponds to the reduction of Einstein's gravitational equations to a set of four gravito-magnetic equations (see, e. g., Peng and Wang [6], Biemond [7] and de Matos and Becker [8]). In our discussion attention will be paid to this remarkable mirror character between electromagnetic and gravitational phenomena.

In addition, the so-called Wilson-Blackett law can be derived from the gravito-magnetic equations, if the "magnetic-type" gravitational field **B** is identified as a magnetic induction field. The Wilson-Blackett law approximately predicts the observed magnetic fields of rotating bodies ranging from rotating metallic cylinders in the laboratory, moons, planets, stars, up to the Galaxy in an amazing way. Therefore, the gravito-magnetic theory may explain the origin of the magnetic fields of celestial bodies as a bonus (see [7] for an extended discussion).

## 2. New "Einstein" equations

The following postulated equations, partly analogous to Einstein's gravitational equations, will be the starting point of the present approach (compare with [2 (§ 95, pp. 330–332), 4 (ch. 17), 9 (ch. 7)])

$$R_{ij} - 1/2\, g_{ij}R = -2\pi c^{-4} q_0 (m_0)^{-1} Q_{ij} = -2\pi c^{-4} q_s Q_{ij} \qquad (4)$$

In these equations the tensor components $R_{ij}$ and $R$ have their usual mathematical form. The quantities $q_0$ and $m_0$ denote the charge and the mass in the same volume element $dV$ and the specific charge $q_s$ is defined by $q_s \equiv q_0(m_0)^{-1}$. The energy-momentum tensor $T_{ij}$ in Einstein's equations has, however, been replaced by an analogous current density tensor $Q_{ij}$ containing a rest charge density $\rho_0$ instead of a mass density. Note that the right hand side of (4) lacks the gravitational constant $G$ and differs from Einstein's equations by a factor $-1/4$ and the specific charge $q_s \equiv q_0(m_0)^{-1}$. The last factor reflects that charge is associated with mass, as will become more clear in the discussion of the special relativistic electric energy and the Lorentz force below. It is noticed, that no electromagnetic field contribution has been included in the right hand side of (4) (see, e. g., Landau and Lifshitz [2 (§ 33)] for that contribution). Moreover, no term with the so-called "cosmological constant" has been added in (4) (see, e. g, [4 (§ 17.3)] for a discussion of that term in Einstein's equations). In this study we shall mainly deal with the current density term $Q_{ij}$ containing a charge density $\rho_0$.

In this work we will first consider systems consisting of a point charge $Q$ and a charge $q_0 = \rho_0 dV$ (as a rule, $q_0 < Q$) moving with a velocity **v** through the Coulomb potential of charge $Q$. The gravitational effects of mass $m_0$ associated with $q_0$ will be neglected, as well as the gravitational effects of mass $M$ associated with charge $Q$ in a first approximation. The scalar value of the radius vector **R** from charge $Q$ to charge $q_0$ will be denoted by $R$, whereas the position vector from charge $q_0$ to a field point $F$ will be denoted by $\mathbf{r}$. When $r \ll R$, the Coulomb potential from charge $Q$ working on charge $q_0$ and the Newtonian potential from mass $M$ (if present) working on mass $m_0$ are approximately constant (static) at the field point $F$.

The components of the symmetric metric tensor may be written as

$$g_{ij} = \eta_{ij} + h_{ij}, \qquad (5)$$

where $\eta_{ij} = (1, -1, -1, -1)$. In the applied weak field approximation the metric components $g_{ij}$ are close to the Minkowski metric components $\eta_{ij}$ (so $|h_{ij}| < 1$).

The following electromagnetic metric written in Cartesian coordinates will be used in this work (Gaussian units are chosen throughout this paper)

$$c^2 d\tau^2 = g_{00} c^2 dt^2 + 2 g_{0\alpha} c dt dx^\alpha + g_{\alpha\beta} dx^\alpha dx^\beta, \qquad (6)$$

where $\alpha = x, y,$ or $z$ and so on. Moreover, for $g_{ij}$ we shall use the tensor components

$$g_{00} = 1 + h_{00} = 1 + 2(1 - v^2/c^2)^{1/2} c^{-2} q_s \phi + (c^{-2} q_s \phi)^2 \cong 1 + 2(1 - v^2/c^2)^{1/2} c^{-2} q_s \phi \cong (g^{00})^{-1} \quad (7)$$

$$g_{0\alpha} = h_{0\alpha} = -(1 - v^2/c^2)^{1/2} c^{-2} q_s A^\alpha - c^{-2} q_s (c^{-2} q_s \phi) A^\alpha \cong -(1 - v^2/c^2)^{1/2} c^{-2} q_s A^\alpha \cong (g^{0\alpha})^{-1} \quad (8)$$

$$g_{\alpha\beta} = -\delta_{\alpha\beta} + (c^{-2} q_s)^2 A^\alpha A^\beta \cong -\delta_{\alpha\beta} \cong (g^{\alpha\beta})^{-1}, \qquad (9)$$

where $\phi = Q/R$ is the Coulomb potential from charge $Q$ working on charge $q_0$ and $A^\alpha$ denote the components of the electromagnetic vector potential $\mathbf{A}$. Potential $\mathbf{A}$ may be due to translation or rotation of $Q$ (or may be due to an externally applied magnetic field). As an example, the dimensonless quantity $c^{-2} q_s \phi$ in (7) and (8) for the electron in the hydrogen atom equals to $e^2/(m_0 c^2 r_0) = 5.33 \times 10^{-5}$ ($e$ is the charge of the electron, $m_0$ its mass (here) and $r_0$ is Bohr's radius). The full metric tensor components (7)–(9) have been calculated from the following electromagnetic Lagrangian (see, e. g., [2 (p. 45)], Goldstein[10])

$$L = -m_0 c^2 \frac{d\tau}{dt} = -(1 - v^2/c^2)^{1/2} m_0 c^2 - q_0 \phi + c^{-1} q_0 \mathbf{A} \cdot \mathbf{v} \qquad (10)$$

Note that in many cases $|c^{-1} q_0 \mathbf{A} \cdot \mathbf{v}| < |q_0 \phi| < m_0 c^2$, so that the approximations made in (7)–(9) can be justified. Therefore, in the sequel of this work the components $g_{0\alpha}$ and $g^{0\alpha}$ will be neglected in appropiate cases. On microscopic level, however, for example, on the level of the substructure of the electron, where the rms $|c^{-1} q_0 \mathbf{A} \cdot \mathbf{v}|$, $|q_0 \phi|$ and $m_0 c^2$ may be comparable in magnitude, the full expressions (7)-(9) may be required. Note that no term containing electromagnetic fields has been included in the right hand side of (10) (compare with [2 (p. 45)]). Finally, it is noticed that taking $d\tau = 0$ for photons (as usually is assumed) one may deduce a light velocity $v = c$ from (10) assuming $q_0$ takes zero value for photons.

Equations (4) will now be evaluated by introducing a new, mixed tensor $\psi^j_i$ defined as

$$\psi^j_i \equiv h^j_i - 1/2 \, \delta^j_i h, \qquad (11)$$

where $h^j_i \equiv g^{jk} h_{ki} \cong g^{jj} h_{ji}$ and $h \equiv h^j_j$ ($j$ summed), and by imposing the four conditions

$$\frac{\partial \psi^j_i}{\partial x^j} = 0 \qquad (12)$$

Here $x^j$ is a four vector: $x^j \equiv (x^0, x, y, z) = (ct, \mathbf{r})$.

For $i = 0$ insertion of (5), (11) and (12) into (4) yields the approximate expression

$$g^{kk}\frac{\partial^2 \psi^j_0}{\partial x^k \partial x^k} = 4\pi c^{-4} q_s g_{00} Q^{0j} \qquad \text{(k summed from 0–3)} \qquad (13)$$

This result is more general than the corresponding linearized Einstein equations given in, e.g., [2 (pp. 330–332), 4 (ch. 18), 6 (ch. 2), 8 (ch. 10)], where $g^{kk}$ and $g_{00}$ are approximated by $\eta^{kk}$ and $\eta_{00}$, respectively. Choosing $r \ll R$ in most of this work the tensor components $g^{kk}$ and $g_{00}$ may assumed to be approximately constant.

The current density tensor $Q^{0j}$ in (13) can be written as (compare with [2 pp. 82, 256, 338)])

$$Q^{0j} = c(g_{00})^{-1/2} \frac{dx^j}{dt} \frac{dt}{d\tau} \rho_0, \qquad (14)$$

where $\rho_0$ is the rest charge density. Introduction of (14) into (13) then yields

$$g^{kk}\frac{\partial^2 \psi^j_0}{\partial x^k \partial x^k} = 4\pi c^{-3} q_s (g_{00})^{1/2} \frac{dx^j}{dt} \frac{dt}{d\tau} \rho_0, \qquad \text{(k summed from 0–3)} \qquad (15)$$

It is noticed that apart from the four equations of (15) following from (4) for $i = 0$ another set of twelve equations follow from (4) for $i \neq j$. The right hand sides of the latter equations contain the velocity up to second order (product of $dx^i/dt$ and $dx^j/dt$) instead of first order ($dx^j/dt$) in (15). In this work we will not consider these second order contributions further.

One now can define the following four vector $A^j = (\varphi, A^\alpha)$ with the scalar component by

$$\varphi \equiv c^2 (q_s)^{-1} \psi^0_0 \cong 1/2\, c^2 (q_s)^{-1} (g_{00})^{-1} h_{00} \qquad (16)$$

and the vector components $A^\alpha$ of vector $\mathbf{A}$ by

$$A^\alpha \equiv c^2 (q_s)^{-1} \psi^\alpha_0 \cong c^2 (q_s)^{-1} (g_{\alpha\alpha})^{-1} h_{\alpha 0} \qquad (\alpha\ \text{not summed}) \qquad (17)$$

In (16) and (17) the approximated right hand sides correspond to the approximated equations (7)–(9). At low velocity and for a charge $Q = q_0$ placed in the field point $F$ it can be seen from (7) and (8) that the Coulomb potential $\phi$ of charge $Q = q_0$ reduces to $\phi = q_0/r$ and corresponds with $\varphi$ of (16), whereas $A^\alpha$ of (8) and (9) corresponds with $A^\alpha$ of (17) under the same conditions. Thus, the seemingly arbitrary definitions for $\varphi$ and $A^\alpha$ in (16) and (17) may be chosen in such a way. It is noticed that related definitions of $\varphi$ and $A^\alpha$ have been used in the reduction of Einstein's equations leading to a set of four gravito-magnetic equations (see, e.g., [6–8]).

Since the components $g_{ii}$ are approximately constant for $r \ll R$, calculation from equation (12) for $i = 0$ utilizing (16) and (17) yields the following generalized Lorentz condition

$$\nabla \cdot \mathbf{A} + c^{-1} \frac{\partial \varphi}{\partial t} = 0, \qquad (18)$$

where $\nabla \equiv (\frac{\partial}{\partial x}, \frac{\partial}{\partial y}, \frac{\partial}{\partial z})$ is the Laplace operator.

Combination of (15)–(18) leads to two second order differential equations, one for the scalar potential $\varphi$ and one for the vector potential $\mathbf{A}$, respectively

$$\nabla^2\varphi - (c')^{-2}\frac{\partial^2\varphi}{\partial t^2} = 4\pi(g_{00})^{1/2}g_{\alpha\alpha}\frac{dt}{d\tau}\rho_0 \qquad (\alpha \text{ not summed}) \qquad (19)$$

$$\nabla^2\mathbf{A} - (c')^{-2}\frac{\partial^2\mathbf{A}}{\partial t^2} = 4\pi c^{-1}(g_{00})^{1/2}g_{\alpha\alpha}\frac{dt}{d\tau}\rho_0\mathbf{v}, \qquad (\alpha \text{ not summed}) \qquad (20)$$

where $\mathbf{v} \equiv (v^x, v^y, v^z)$ and $\alpha = x, y$ or $z$. For convenience sake, it has been assumed that $g_{xx} = g_{yy} = g_{zz} = g_{\alpha\alpha}$ ($\alpha$ not summed). Note that in the full expression (9) the components $g_{xx}$, $g_{yy}$ and $g_{zz}$ may be different. The velocity $c'$ in (19 and (20) is given by

$$c' = (g_{00})^{1/2}(-g_{\alpha\alpha})^{-1/2}c \qquad (\alpha \text{ not summed}) \qquad (21)$$

Note that for a plane light wave $q_0$ may be assumed to be zero. In that case, utilizing (7) and (9), $c'$ in (21) reduces to $c$. So, for a plane wave, both the scalar potential $\varphi$ and the vector potential $\mathbf{A}$ propagate with light velocity $c$ in the presence of a charge $Q$.

### 3. Generalized electromagnetic field tensor

In this section generalized electromagnetic field tensors $F_{ij}$ and $F^{ij}$ will be introduced (compare with, e.g., Landau and Lifshitz [2(§90)] and Van Bladel [3, ch. 9]). All these authors assumed the presence of a big mass $M$, but we will first consider the case where a big charge $Q$ is present. The tensor components of $F_{ij}$ will be defined as

$$F_{ij} \equiv \frac{\partial A_j}{\partial x^i} - \frac{\partial A_i}{\partial x^j}, \qquad (22)$$

where the covariant vector $A_i$ is connected to the contravariant vector $A^j = (\varphi, A^\alpha)$ from (16) and (17) by the relation

$$A_i = g_{ij}A^j \cong g_{ii}A^i \qquad (23)$$

The non-diagonal components of the metric tensor $g_{ij}$ are neglected in (23), whereas the diagonal components $g_{ii}$ are taken constant for $r \ll R$. Likewise, the covariant tensor $F_{ij}$ is connected to the contravariant tensor $F^{ij}$ by

$$F_{ij} = g_{ik}g_{jl}F^{kl} \cong g_{ii}g_{jj}F^{ij}, \qquad (24)$$

The components of the generalized electric field $\mathbf{E}$ and the generalized magnetic induction field $\mathbf{B}$ are then defined in the usual way by, respectively

$$E_x \equiv F_{0x}, E_y \equiv F_{0y} \text{ and } E_z \equiv F_{0z} \qquad (25)$$

$$B_x \equiv F_{zy}, B_y \equiv F_{xz} \text{ and } B_z \equiv F_{yx} \qquad (26)$$

Combination of (22)–(25) then yields for the field $\mathbf{E}$

$$\mathbf{E} = -g_{00}\nabla\varphi + c^{-1}g_{\alpha\alpha}\frac{\partial \mathbf{A}}{\partial t}, \qquad (\alpha \text{ not summed}) \qquad (27)$$

whereas from combination of (22)–(24) and (26) follows

$$\mathbf{B} = -g_{\alpha\alpha}\nabla\times\mathbf{A} \qquad (\alpha \text{ not summed}) \qquad (28)$$

It is noticed that the fields **E** of (27) and **B** of (28) depend on the tensor components $g_{00}$ and/or $g_{\alpha\alpha}$, the definitions (22), (25) and (26) and the transformations (23) and (24). Likewise, the components of the generalized electric displacement field **D** and the generalized magnetic field **H** may be defined in the usual way by, respectively

$$D^x \equiv F^{x0}, D^y \equiv F^{y0} \text{ and } D^z \equiv F^{z0}, \tag{29}$$

$$H^x \equiv F^{zy}, H^y \equiv F^{xz} \text{ and } H^z \equiv F^{yx} \tag{30}$$

On now may introduce the usual "constitutive" equations

$$\mathbf{D} = \varepsilon_0 \mathbf{E} \text{ and } \mathbf{B} = \mu_0 \mathbf{H}, \tag{31}$$

where, $\varepsilon_0$ and $\mu_0$ are the vacuum permittivity and the vacuum permeability, respectively, in the presence of a charge $Q$. Combination of (24), (25) and (29) using the approximated equations (7) and (9) then yields for $\varepsilon_0$

$$\varepsilon_0 = -(g_{00} g_{\alpha\alpha})^{-1} \cong 1 - 2(1 - v^2/c^2)^{1/2} c^{-2} q_s \phi \quad (\alpha \text{ not summed}) \tag{32}$$

This relation reveals a dependence of $\varepsilon_0$ on the electric potential $\phi$ of the vacuum itself and implies a polarization $\mathbf{P} \equiv (4\pi)^{-1}(\varepsilon_0 - 1)\mathbf{E} = \chi \mathbf{E}$ of the vacuum. The sign of the dielectric susceptibilty $\chi$ depends on the signs of the charges $q_0$ and $Q$: for $\chi > 0$ the vacuum behaves like a "dia-electric" medium and for $\chi < 0$ like a "para-electric" medium. Combination of (24), (26) and (30) using the approximated equation (9) yields for $\mu_0$

$$\mu_0 = (g_{\alpha\alpha})^2 \cong 1 \quad (\alpha \text{ not summed}) \tag{33}$$

Combination of (21), (32) and (33) yields the relation

$$\varepsilon_0 \mu_0 = -(g_{00})^{-1} g_{\alpha\alpha} = (c/c')^2 \quad (\alpha \text{ not summed}) \tag{34}$$

It is noticed that by using equations (31)–(33) Maxwell equations can be formulated, analogous to the Maxwell equations for electromagnetic fields in material media, as will be shown in sections 4 and 7.

## 4. Special relativistic Maxwell equations

From (18)–(21), (27), (28) and (31)–(33), by making use of standard vector analysis, one then can derive the following set of four special relativistic Maxwell equations for sources in vacuum

$$\nabla \times \mathbf{H} = 4\pi c^{-1} (g_{00})^{1/2} \frac{dt}{d\tau} \rho_0 \mathbf{v} + c^{-1} \frac{\partial \mathbf{D}}{\partial t} \tag{35}$$

$$\nabla \cdot \mathbf{D} = 4\pi (g_{00})^{1/2} \frac{dt}{d\tau} \rho_0 \tag{36}$$

$$\nabla \times \mathbf{E} = -c^{-1} \frac{\partial \mathbf{B}}{\partial t} \tag{37}$$

$$\nabla \cdot \mathbf{B} = 0 \tag{38}$$

Note that evaluation of (35)–(38) for a plain light wave ($\rho_0 = 0$) yields the common isotropic light velocity $c$. See comment after (21), too.

It is also noticed that the following special relativistic charge density $\rho$ may be separated off from (35) and (36)

$$\rho = (g_{00})^{1/2} \frac{dt}{d\tau} \rho_0 \tag{39}$$

The quantity $dt/d\tau$ in (39) can be approximated by series expansion of (10)

$$\frac{dt}{d\tau} \cong (1 - v^2/c^2)^{-1/2} - (1 - v^2/c^2)^{-1} c^{-2} q_s \phi, \tag{40}$$

where the approximations $|c^{-1} q_0 \mathbf{A} \cdot \mathbf{v}| < |q_0 \phi| < m_0 c^2$ have been used. Introduction of the approximate equations (7) and (40) into (39) yields for $v < c$

$$\rho \cong (1 - v^2/c^2)^{-1/2} (1 - v^2 c^{-4} q_s \phi) \rho_0 \tag{41}$$

When $Q$ is absent, $\rho$ of (41) reduces to expression (1), a result earlier obtained by Jefimenko [1]. He found that the Maxwell equations in which the rest charge density $\rho_0$ was replaced by the charge density $\rho$ of (1) remained invariant (retained their form) under a special relativistic transformation. For example, his equation (1) (corresponding to our equation (36) with $\rho$ replaced by $\rho_0$) did not appear to be invariant under a special relativistic transformation, but $\rho = (1 - v^2/c^2)^{-1/2} \rho_0$ did.

Comparison with the corresponding Maxwell equations obtained by Landau and Lifshitz [2(§90)] shows, that their equations contain the rest charge density $\rho_0$ instead of a special relativistic charge density like $\rho$ of (1). The crucial missing factor $(1 - v^2/c^2)^{-1/2}$ placed me on the track in finding the equations (4). Since the equations of the authors of [2] lack this factor, their derivation does not seem satisfying. On the other hand, the obtained special relativistic equations (35)–(38) may be a first indication of the validity of the postulated equation (4).

Finally, by combining (35) and (36) in first order (see (41) and added comment) the following relation can be calculated

$$\nabla \cdot (\rho \mathbf{v}) + \frac{\partial \rho}{\partial t} = 0 \tag{42}$$

This expression embodies the conservation law of the charge density $\rho$ given by (1).

## 5. Non-relativistic case

At a non-relativistic velocity $\mathbf{v}$ ($v \ll c$) and in absence of an additional charge $Q$ (19) and (20) reduce to the well known second order differential equations for $\varphi$ and $\mathbf{A}$ (both containing the rest charge density $\rho_0$), respectively, whereas (18) reduces to the familiar Lorentz condition (see, e.g., [2(p. 158)]. In addition, in the non-relativistic case equations (35)–(38) reduce to the so-called vacuum Maxwell equations (In fact, Maxwell (1864) gave a more extended system of equations. Heaviside reformulated them in a more familiar form (see, e.g., Nahin [11] for a historical exposition). Therefore, we have consequently denoted the set of four equations as Maxwell equations instead of Maxwell's equations). Note that in the non-relativisic case $\varepsilon_0$ from (32) and $\mu_0$ from (33) reduce to unity value. From (31) then follows that in that case $\mathbf{D}$ coincides with $\mathbf{E}$ and $\mathbf{B}$ with $\mathbf{H}$, respectively.

It is noticed that a comparable reduction of Einstein's equations leading to a set of four gravito-magnetic equations, almost analogous to the Maxwell equations has been found earlier (see, e.g., [6–8]).

## 6. Special relativistic electric energy and Lorentz force

Starting from the metric components of (6) the conservation law of electric energy and the equations of motion can be calculated (see, e.g., Landau and Lifshitz [2(§87)]) from the following full expression

$$\frac{du_i}{ds} = 1/2 \frac{\partial g_{kl}}{\partial x^i} u^k u^l, \qquad (43)$$

where $ds = cd\tau$ and $dx_i/ds = u_i = g_{ij}u^j$. If appropriate expressions for the metric components are substituted into (43), (43) will apply both to the electromagnetic case and the gravitational case. Evaluation of (43) (see, e.g., de Matos and Becker [8]) yields

$$\frac{d\tau}{dt}\left\{\frac{d}{dt}\left(g_{ij}\frac{dx^j}{dt}\frac{dt}{d\tau}\right)\right\} = 1/2 \frac{\partial g_{kl}}{\partial x^i}\frac{dx^k}{dt}\frac{dx^l}{dt} \qquad (44)$$

For $i = 0$ the conservation law of energy follows from (44)

$$\frac{d\tau}{dt}\left[\frac{d}{dt}\left\{\left(cg_{00} + g_{0\alpha}\frac{dx^\alpha}{dt}\right)\frac{dt}{d\tau}\right\}\right] = 1/2\, c\frac{\partial g_{00}}{\partial t} + \frac{\partial g_{0\alpha}}{\partial t}\frac{dx^\alpha}{dt} + 1/2\, c^{-1}\frac{\partial g_{\alpha\beta}}{\partial t}\frac{dx^\alpha}{dt}\frac{dx^\beta}{dt} \qquad (45)$$

Substitution of the approximated equations (7)–(9) and (40) followed by evaluation yields for $v < c$

$$\frac{dH}{dt} = \frac{d}{dt}\left\{(1 - v^2/c^2)^{-1/2}m_0c^2 + q_0\phi\right\} \cong 0, \qquad (46)$$

where the conserved quantity $H$ is the well known total electric energy of the system (see, e.g., [2(p.46)], Goldstein [10]).

For $i = \alpha$ ($\alpha = x, y,$ or $z$) the equations of motion follow from (44)

$$\frac{d\tau}{dt}\left[\frac{d}{dt}\left\{\left(cg_{\alpha 0} + g_{\alpha\beta}\frac{dx^\beta}{dt}\right)\frac{dt}{d\tau}\right\}\right] = 1/2\, c^2\frac{\partial g_{00}}{\partial x^\alpha} + c\frac{\partial g_{0\beta}}{\partial x^\alpha}\frac{dx^\beta}{dt} + 1/2\frac{\partial g_{\beta\gamma}}{\partial x^\alpha}\frac{dx^\beta}{dt}\frac{dx^\gamma}{dt} \qquad (47)$$

Substitution of the approximated equations (7)–(9), (16), (17), (27), (28) and (40) followed by evaluation yields for $v < c$

$$\frac{d}{dt}\left\{(1 - v^2/c^2)^{-1/2}m_0\mathbf{v}\right\} = q_0\mathbf{E} + c^{-1}q_0\mathbf{v}\times\mathbf{B} \qquad (48)$$

This result corresponds to the usual expression for the special relativistic Lorentz force (see, e.g., Goldstein [10]), but deviations may occur when $|c^{-1}q_0\mathbf{A}\cdot\mathbf{v}| \cong |q_0\phi| \cong m_0c^2$ or when $v \cong c$. Note that both equation (46) and (48) contain a charge $q_0$ as well as a rest mass $m_0$. In the corresponding gravitational equations following from Einstein's equations mass can occur without charge. Therefore, (46) and (48) might reflect that mass is a more fundamental quantity than charge. It is noticed that equations (46) and (48) are generally accepted to be valid. Thus, (46) and (48) may be a second indication of the validity of the postulated equations (4). Moreover, the introduction of the electromagnetic metric components (7)–(9) into (43) may be justified. Components $g_{ij}$ have, however, to be solutions of equations (4), too. Higher order terms in presented components of (7)–(9) may not fulfil this requirement.

## 7. General relativistic Maxwell equations

In this section we will now consider systems consisting of a big mass $M$ and a charge $q_0 = \rho_0 dV$ moving with a velocity $\mathbf{v}$ through the gravitational potential of $M$. No additional charge $Q$ or vector potential $\mathbf{A}$ will be assumed to be present. In that case the gravitational contributions from $M$ have to be introduced into the metric tensor (6). (The gravitational effects of mass $m_0$ associated with $q_0$ will be neglected.) The following metric may then be chosen (see, e.g., [2(pp. 337–338 and p. 304)])

$$c^2 d\tau^2 \cong (1 - 2GM/c^2R)c^2 dt^2 - (1 + 2GM/c^2R)(dx^2 + dy^2 + dz^2) \qquad (49)$$

So, all metric components $g_{ij}$ and $g^{ij}$ for $i \neq j$ are neglected, whereas in the diagonal components $g_{ii}$ and $g^{ii}$ only first order terms in $GM/c^2R$ are considered [2(pp. 337–338, see also, p. 304)]. The following non-zero tensor components $g_{ij}$ are left

$$g_{00} \cong (g^{00})^{-1} \cong 1 - 2GM/c^2R \qquad (50)$$

$$g_{\alpha\alpha} \cong (g^{\alpha\alpha})^{-1} \cong -(1 + 2GM/c^2R), \qquad (\alpha \text{ not summed}) \qquad (51)$$

where $\alpha = x, y,$ or $z$. It is noticed that taking again $d\tau = 0$ for photons the following light velocity $c'$ can be deduced from (49)–(51) (compare with (2), (3) and (21))

$$c' = (g_{00})^{1/2}(-g_{\alpha\alpha})^{-1/2} c \cong 1 - 2GM/c^2R \qquad (\alpha \text{ not summed}) \qquad (52)$$

It is noticed that both metric components (7)–(9) and (50)–(51) apply to equation (15), the definitions (16) and (17) and the relations (18)–(31). Utilizing (50) and (51), evaluation of (32) and (33) yields in the gravitational case, respectively

$$\varepsilon_0 = -(g_{00} g_{\alpha\alpha})^{-1} \cong 1 \qquad (\alpha \text{ not summed}) \qquad (53)$$

$$\mu_0 = (g_{\alpha\alpha})^2 \cong 1 + 4GM/c^2R \qquad (\alpha \text{ not summed}) \qquad (54)$$

This relation implies a magnetization $\mathbf{M} \equiv (4\pi)^{-1}(\mu_0 - 1)\mathbf{H} = \chi\mathbf{H}$. So, in the presence of a strong gravitational field the vacuum behaves like a diamagnetic medium with $\chi > 0$.

Results (53) and (54) may be compared with $\varepsilon_0 = \mu_0 \cong 1 + 2GM/c^2R$, obtained by Puthoff [12] from a polarizable-vacuum representation of general relativity, in combination with the adopted constancy of the fine structure constant. In both approaches the same velocity of light $c'$ of (52) is obtained. It is stressed that our results for $\varepsilon_0$ and $\mu_0$ stem from the definitions of $\mathbf{E}$, $\mathbf{B}$, $\mathbf{D}$ and $\mathbf{H}$ and from the used transformation for the electromagnetic tensor $F_{ij}$ (see section 3).

In the gravitational case combination of (18)–(21), (27), (28), (31), (53) and (54) yields the following set of general relativistic Maxwell equations for sources in vacuum

$$\nabla \times \mathbf{H} = 4\pi c^{-1} \rho \mathbf{v} + c^{-1} \frac{\partial \mathbf{D}}{\partial t} \qquad (55)$$

$$\nabla \cdot \mathbf{D} = 4\pi\rho \qquad (56)$$

$$\nabla \times \mathbf{E} = -c^{-1} \frac{\partial \mathbf{B}}{\partial t} \qquad (57)$$

$$\nabla \cdot \mathbf{B} = 0 \tag{58}$$

Note that in the presence of a gravitational field the charge density $\rho$ of (39) formally does not change in (55) and (56). Series expansion of $dt/d\tau$ from (49) approximately yields

$$\frac{dt}{d\tau} \cong (1 - v^2/c^2)^{-1/2}(1 + GM/c^2R) \tag{59}$$

Using (50) and (59) $\rho$ in (55) and (56) can then approximated by

$$\rho = (g_{00})^{1/2}\frac{dt}{d\tau}\rho_0 \cong (1 - v^2/c^2)^{-1/2}\rho_0 \tag{60}$$

Note that approximate expression of $\rho$ does not contain a first order term $GM/c^2R$.

In order to derive equations (55)–(58) it is also possible to follow the method of Landau and Lifshitz [2 (§ 90)] and Van Bladel [3 (ch. 9)] and start from two relations in terms of $F_{ij}$ and $F^{ij}$, respectively

$$\frac{\partial F_{ij}}{\partial x^k} + \frac{\partial F_{ki}}{\partial x^j} + \frac{\partial F_{jk}}{\partial x^i} = 0 \tag{61}$$

Using (25) and (26) from (61) the second pair of generalized Maxwell equations (57) and (58) can then be calculated. In a similar way using (29) and (30) the first pair of generalized Maxwell equations (55) and (56) can be obtained from

$$\frac{\partial F^{ij}}{\partial x^j} = -4\pi c^{-1}\rho\frac{dx^i}{dt}, \tag{62}$$

where $\rho$ has been given by the full expression of $\rho$ in (60). It is noticed, however, that in [2 (§ 90)] the corresponding expression for (62) and thus for (55) and (56) contain the rest charge density $\rho_0$ instead of generalized charge density $\rho$ of (60).

In addition, for a plane light wave the isotropic light velocity $c'$ of (52) or (3) is obtained from (55)–(58). From the general relativistic Maxwell equations given in [2 (§ 90), 3 (ch. 9)] the light velocity $c'$ of (2) can be deduced. The latter result also placed me on the track in finding equations (4). Result (2) is unsatisfactory in view of (52), which is compatible with (49) and with the observed effects of the gravitational red-shift and the bending of light in the presence of a gravitational field. Therefore, our result (52) obtained from (55)–(58) may be regarded as a third indication for the validity of (4). Moreover, the choice of the gravitational metric components (50) and (51) applied in the electromagnetic case, for example, in equations (55) and (56) seems to be justified.

## 8. Discussion

The new, postulated basic equations (4) containing a charge density in combination with a scalar potential and vector potential are the basis of a long list of formulas that describe electromagnetic phenomena. A number of examples of such formulas are: generalized Maxwell equations applying at relativistic velocities and in the presence of an additional charge $Q$ (equations (35)–(38)), or an additional big mass $M$ (equations (55)–(58)) and the special relativistic formulas for the electric energy from (46) and the Lorentz force in (48), respectively. It is noticed, that the electromagnetic field contribution has not been included in the right hand side of (4) (see, e. g., Landau and Lifshitz [2 (§ 33)] for that contribution). Moreover, no term with "cosmological constant" has been added in (4) (see, e. g, [4 (§ 17.3)] for a discussion of the latter term in Einstein's equations). In this study we mainly concentrated us on the current density term $Q_{ij}$ containing the charge density $\rho_0$.

Einstein's gravitational equations contaning a mass density are also the basis of a long list of formulas that describe gravitational phenomena. When the gravito-magnetic approach is followed (see, e. g., Peng and Wang [6], Biemond [7] and de Matos and Becker [8]), the following results were calculated from the linearized Einstein equations: a set of four gravito-magnetic equations and the gravito-magnetic force. These results may, however, be generalized in order to include a static gravitational potential. Compare with comment after (13) and with section 7.

In addition, it has been demonstrated by Biemond [7] that in the stationary case the gravito-magnetic equations may lead to a bonus: the so-called Wilson-Blackett law. This law approximately predicts the observed magnetic fields of rotating bodies ranging from rotating metallic cylinders in the laboratory, moons, planets, stars, up to the Galaxy in an amazing way (see [7] for an extended discussion).

We now consider the results obtained from the postulated equations (4) more in detail. At relativistic velocity and in the presence of an additional big charge $Q$ equations (4) reduce to the special relativistic equations (35)–(38). In that case the vacuum permittivity $\varepsilon_0$ of (32) may deviate from unity value. When charge $Q$ is absent, equations (35)–(38) reduce to Maxwell equations (see equations (39)–(41)), first obtained by Jefimenko [1]. The latter equations were found to be invariant (retained their form) under a special relativistic transformation. Moreover, the expected special relativistic electric energy from (46) and Lorentz force in (48) are found, when the approximate, electromagnetic metric components $g_{ij}$ from (7)–(9) are introduced.

One expects there exist generalized vacuum Maxwell equations for a charge source moving at a relativistic velocity in the presence of a static gravitational potential. From the postulated equations (4) such a set of four differential equations (equations (55)–(58)) is obtained. Contrary to the result (2) of Landau and Lifshitz [2, (§90)] and Van Bladel [3 (ch. 9)], the expected isotropic light velocity $c'$ of (52) or (3) has been obtained from these equations. Moreover, in the presence of a big additional mass $M$ the vacuum permittivity $\mu_0$ of (54) deviates from unity value.

In this paper we have restricted us to the isotropic metric (49), but an exact solution of Einstein's equations, the Schwarzschild metric (see, e. g., Landau and Lifshitz [2, (§ 100)], is anisotropic

$$c^2 d\tau^2 \cong (1-2GM/c^2R)c^2dt^2 - (1-2GM/c^2R)^{-1}dR^2 - R^2d\theta^2 - R^2\sin^2\theta d\varphi^2 \qquad (63)$$

So, for $d\tau = 0$ an anisotropic velocity of light follows from (63): the velocity of light perpendicular and parallel to the direction of the gravitational field equals to the result of (2) and (3), respectively. A more extended treatment should account for this effect.

The full consequences of the postulated basic equations (4) have not yet been considered (mainly because first order approximations have been made in this paper), but they may become manifest at a relativistic velocity and in the strong electric field limit (and in the strong gravitational field limit). Therefore, equations (4) may appear to be important on the microscopic level, for example, on the level of the substructure of leptons and quarks. Harari [13], e. g., proposed two basic building blocks: the T-rishon with charge 1/3 e (e is charge of the positron) and the V-rishon without charge. Thus, equation (41) or (1) may become manifest in the description of the interaction mechanism of T-rishons in the positron, for instance.

It is noticed that the right side of equations (4) may be added to Einstein's gravitational equations yielding

$$R_{ij} - 1/2\, g_{ij}R = 8\pi c^{-4}GT_{ij} - 4\pi c^{-4}q_s Q_{ij} \qquad (64)$$

Neither the usual electromagnetic field contribution in the tensor $T_{ij}$ has been included in (64) (compare with, e. g., Landau and Lifshitz [2 (§ 33)] for this contribution), nor the term with "cosmological constant" (see, e. g, [4 (§ 17.3)]. As a result, the tensor $T_{ij}$ only contains a mass density, whereas the tensor $Q_{ij}$ contains a charge density, analogous in form to the tensor $T_{ij}$. The metric components $g_{ij}$ may contain charge and mass contributions. If the term $T_{ij}$ may be neglected, equations (4) are left. The metric components $g_{ij}$ may then contain charge or (/and) mass contributions (see sections 4 and 7).

At the completion of this paper, I found a related treatment of electromagnetism in the presence of a gravitational field from Beach [14]. In this work it was also suggested that metric components may contain charge sources, but the charge source term was not isolated from the electromagnetic field contribution in the energy-momentum tensor $T^{ij}$. Moreover, no explicit expressions for $g_{ij}$, electromagnetic field tensor $F^{ij}$, generalized Maxwell equations and light velocity were given.

Einstein's equations and the new equations (4) and derived formulas form a pair of mirrors reflecting our knowledge of nature. The proposed equations (4) may fill a hole in existing theory. The found analogy between electromagnetic and gravitational quatities in (64) is striking, but incomplete. For example, the full electromagnetic metric components of $g_{ij}$ of (7)–(9) are only partly analogous to the gravitational metric components $g_{ij}$ of (50) and (51).

Summing up, Einstein's equations describing gravitational phenomena are at the top of a theoretical mirror that may be reflected by another mirror with at the top the postulated basic equations (4) describing electromagnetic phenomena. To my knowledge the new, postulated equations (4) are compatible with all known theoretical and experimental evidence. Therefore, presented work may be a step towards further unification of existing physical theories.

## Acknowledgement

I wish to thank my son Pieter for technical realisation and publication of this paper.